\documentclass[manuscript]{emulateapj}
\usepackage[english]{babel}
\usepackage{blindtext}
\usepackage{graphicx}
\usepackage{subfigure}
\usepackage{natbib}
\usepackage{amsmath}
\usepackage{color}
\usepackage{times}

\slugcomment{Accepted for publication in The Astrophysical Journal Letters}
\shorttitle{Blue Straggler Stars Population in NGC\,1261}
\shortauthors{Simunovic, Puzia \& Sills}

\begin{document}

\title{The Blue Straggler Star Population in NGC\,1261: Evidence for a Post-Core-Collapse Bounce State}
\author{Mirko Simunovic$^{1,2}$, Thomas H. Puzia$^{1}$, Alison Sills$^{3}$}
\affil{$^{1}$Institute of Astrophysics, Pontificia Universidad Cat\'olica de Chile, Avenida Vicu\~na Mackenna 4860, 7820436 Macul, Santiago, Chile \\
$^{2}$Astronomisches Rechen-Institut, Zentrum f\"ur Astronomie der Universit\"at Heidelberg, M\"onchhofstr. 12-14, 69120 Heidelberg, Germany \\
$^{3}$Department of Physics \& Astronomy, McMaster University, 1280 Main Street West, Hamilton, ON, L8S 4M1, Canada}
\email{Email: msimunov@astro.puc.cl, tpuzia@astro.puc.cl, asills@mcmaster.ca}

\begin{abstract}
We present a multi-passband photometric study of the Blue Straggler Star (BSS) population in the Galactic globular cluster (GC) NGC\,1261, using available space- and ground-based survey data.~The inner BSS population is found to have two distinct sequences in the color-magnitude diagram, similar to double BSS sequences detected in other GCs. These well defined sequences are presumably linked to single short-lived events such as core collapse, which are expected to boost the formation of BSSs.~In agreement with this, we find a BSS sequence in NGC\,1261 which can be well reproduced individually by a theoretical model prediction of a 2 Gyr old population of stellar collision products, which are expected to form in the denser inner regions during short-lived core contraction phases.~Additionally, we report the occurrence of a group of BSSs with unusually blue colours in the CMD, which are consistent with a corresponding model of a 200 Myr old population of stellar collision products.~The properties of the NGC\,1261 BSS populations, including their spatial distributions, suggest an advanced dynamical evolutionary state of the cluster, but the core of this GC does not show the classical signatures of core-collapse.~We argue these apparent contradictions provide evidence for a post-core-collapse bounce state seen in dynamical simulations of old GCs. \end{abstract}


\section{Introduction}
The study of globular clusters (GCs) offers a rich platform for the validation of stellar evolution theories and the formulation of dynamical evolution scenarios of stellar systems.~In particular, recent discoveries \citep[e.g.][]{fer12,knigge09} have prompted Blue Straggler Stars (BSSs) as key elements of the puzzle, given their tight relation with these two fields.~BSSs are empirically defined as stars that appear bluer and brighter than turn-off stars in the color-magnitude diagram (CMD) of a GC. Although their definite physical nature is still a matter of debate, the most popular formation channels for BSSs are the ones explaining them as products of mass-transfer/mergers in binary stellar systems \citep{mccrea64} and as the results of direct stellar collisions \citep{hillsday76}.~Both are strongly dependent on the local environment's dynamical state, and, therefore, BSSs can be used as tracers of past dynamical events as well as of the current dynamical state, given their strong response to two-body relaxation effects due to their relatively high stellar masses.~\cite{fer12} proposed that the current BSS radial density profile can be used as a ``dynamical clock" to estimate the dynamical age of a GC, and continued to classify a large set of GCs in three different classes according to their dynamical state.~The dynamically oldest category contains, among others, GCs M\,30 and NGC\,362, which have been studied in this context more in detail by \citet[hereafter F09]{fer09} and \citet[hereafter D13]{dal13}.~These studies have found that the BSS population is indeed consistent with dynamically old GCs.~Furthermore, their CMDs reveal well defined double BSS sequences, which the authors claim are the result of single short-lived dynamical events, such as core-collapse.~It is believed that during such core contraction the stellar collision rate would become enhanced, producing the bluer BSS sequence in the CMD, while the boosted binary interaction rate would lead to enhanced Roche-lobe overflows, thereby producing the redder BSS sequence.~It is, therefore, clear that observational properties of BSSs can provide valuable information for the understating of the dynamical evolution of GCs.~Here, we present a study based principally on high-quality HST photometry of the inner BSS population in the poorly studied Galactic GC NGC\,1261, which contains evidence for a very particular dynamical history, making it a similar, yet unique case among other GCs with well defined BSS features, such as M\,30 and NGC\,362.

\section{Data Description}
The inner region photometric catalog comes from the HST/ACS Galactic Globular Cluster Survey \citep{sar07}.~It consists of $\sim$30 min.~exposures in the F606W ($\sim\!V$) and F814W ($\sim\!I$) bands for the central $3.4\arcmin\times3.4\arcmin$ field of NGC\,1261.~The photometry was corrected to account for updated HST/ACS WFC zero points and calibrated in the Vega photometric system.~The catalog provides high quality photometry down to $\sim\!6$ mag below the main-sequence turn-off. Additionally, we performed PSF photometry using the DoPHOT software package \citep{sch93, alo12} on HST/WFC3 data taken with the F336W ($\sim\!U$) band, available from the Hubble Legacy Archive (PI: Piotto, Proposal ID: 13297).~The photometry was calibrated using Stetson standards \citep{ste00} from the NGC\,1261 field. The astrometry was refined with the HST/ACS optical catalog using bright isolated stars, after which we reach a median accuracy of $\sim\!0.002\arcsec$ between the optical and F336W-band catalogs.~We also use wide-field photometry from the catalog published by \cite{kra10}, built from observations at the 1.3-m Warsaw telescope at Las Campanas Observatory, using a set of $UBVI$ filters and a $14\arcmin\times14\arcmin$ field of view. The photometry is calibrated to \cite{ste00} and the median error is $\leqslant$0.04 mag for all filters and colors down to $V=20$ mag.~The complete description of their data reduction and photometric calibration can be found in \cite{kra10}.

\section{The central BSS population in NGC\,1261}
\begin{figure}[t!]
\centering
\includegraphics[width=9cm]{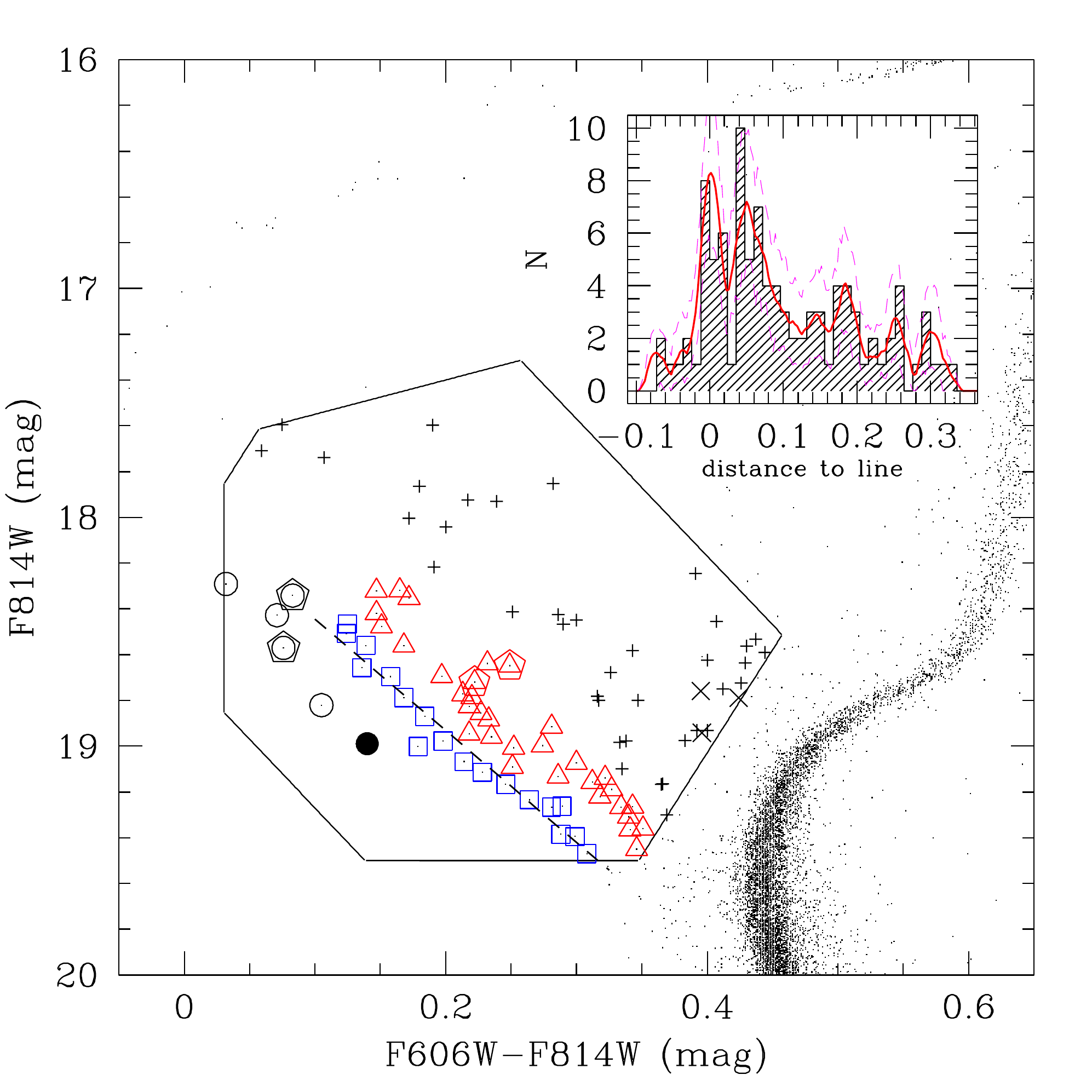}
\caption{\small{(F606W-F814W)\,vs.\,F814W CMD of the NGC\,1261 inner region. All detections come from the final matched HST catalog.~The R-BSS (red triangles), B-BSS (blue squares) and eB-BSS (black circles; filled circle is explained in Fig~\ref{Ucmd}) sub-samples are marked; crosses mark the rest of the BSS sample (except for diagonal crosses, see Fig~\ref{Ucmd}).~Pentagon symbols mark special cases, see Fig~\ref{Ucmd} .~The inset panel shows the distribution of BSS perpendicular distances from the best-fit line to the B-BSS sequence (shown as a dashed line) in mag units.~The solid red line shows a non-parametric Epanech\-nikov-kernel probability density estimate with 90\% confidence limits represented by the dotted pink lines.} }
\label{cmd1}
\end{figure}

BSSs are selected through their position in a CMD. Having three filters, we can use the additional color information to remove contaminants. First, we cross-match the optical ACS catalog with the F336W-band catalog and keep matched sources with separations $<$ 0.02\arcsec\ ($\sim\!0.5$ pix) and reported errors\footnote{Taken from the HST/ACS catalog. More information on the errors is found on the catalog's README file.} $<$ 0.03 mag in the F606W and F814W filter.~This results in $\sim\!25000$ sources in a $\sim\!2.7\arcmin\!\times\!2.7\arcmin$ field centered on NGC\,1261 (see Figure~\ref{cmd1}).~We use a F814W\,$<$\,19.5 mag limit for the BSS selection criteria from \cite{lei11}, who define the BSS region based on magnitude and color cuts in the (F606W-F814W)\,vs.\,F814W CMD, shown by the polygon in Figure~\ref{cmd1}.~All stars inside the region are considered to be BSS candidates, and used in the following analysis (unless removed from the sample; see further down in the text).

\begin{figure}[t!]
\centering
\includegraphics[width=9cm]{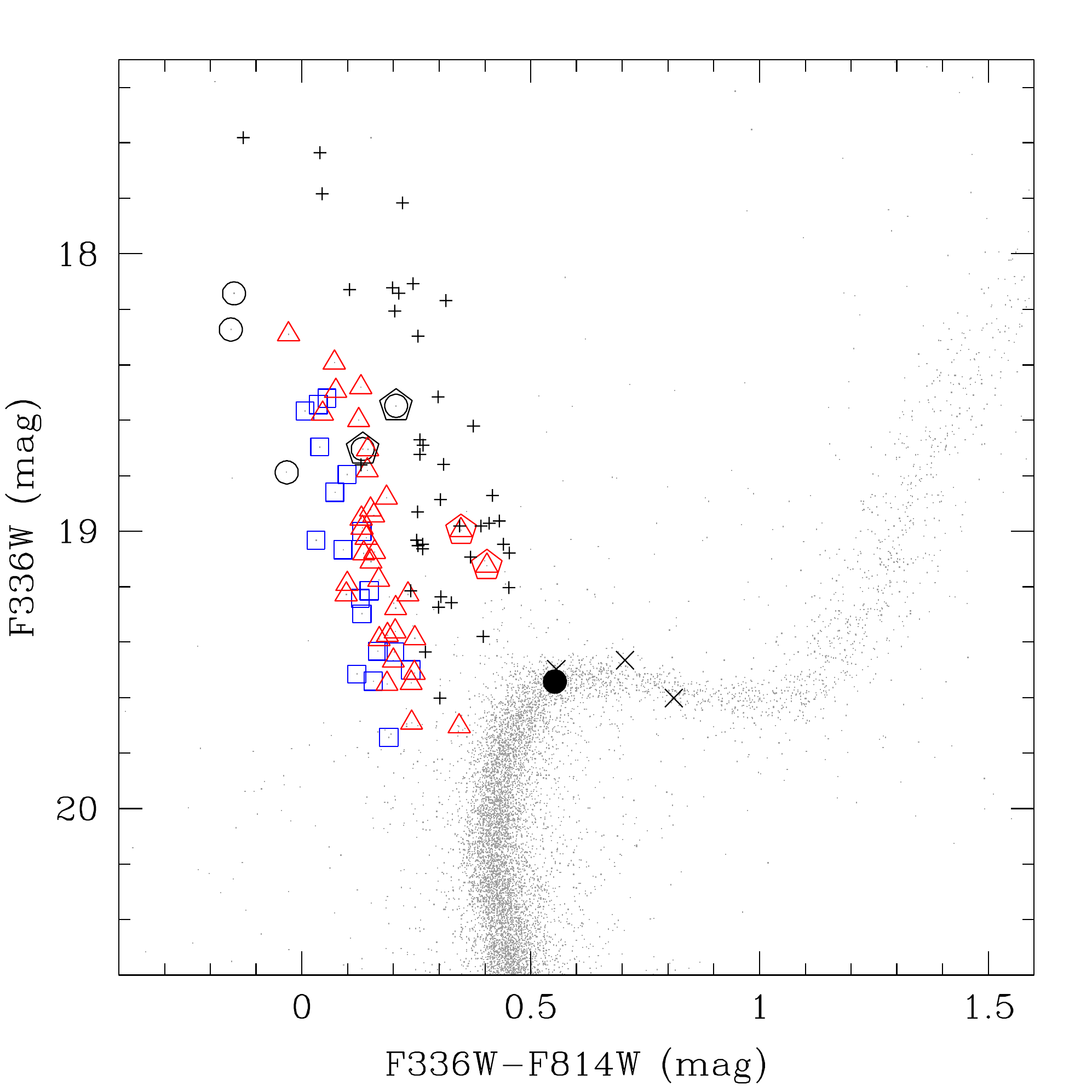}
\caption{\small{(F336W-F814W)\,vs.\,F336W CMD of the NGC\,1261 inner region.~Filled and $\times$-shape symbols indicate BSS candidates that were rejected from our BSS sample.~Pentagon symbols mark BSS candidates that appear slightly faint in the F336W band, relative to their parent subsamples.} }
\label{Ucmd}
\end{figure}

We identify two prominent BSS sequences in the CMD, each containing $\sim\!20$ BSSs (see inset panel).~We label them the blue-sequence-BSSs (B-BSSs; shown as blue squares) and the red-sequence-BSSs (R-BSSs, red triangles).~Similar to M\,30 and NGC\,362\footnote{Note that in those studies, the HST F555W filter is used as a V band proxy, while we use the F606W filter.}, the B-BSS sequence is narrower and better defined, while the R-BSS sequence is dispersed towards redder colours.~We note the appearance of a group of BSSs lying bluer from the B-BSS sequence in the CMD. Since they cannot be directly associated with the B-BSS sequence, as they are clearly separated in the diagram, we choose to label them as extremely-blue-BSSs (eB-BSSs, black circles) and we later discuss their likelihood of being associated to a particular BSS population.~This bluer component has neither been observed in M\,30 nor in NGC\,362, and, if confirmed real, requires a given BSS formation scenario, particular to the history of NGC\,1261, that is not present in the other two GCs.

\begin{figure}[t!]
\centering
\includegraphics[width=9cm]{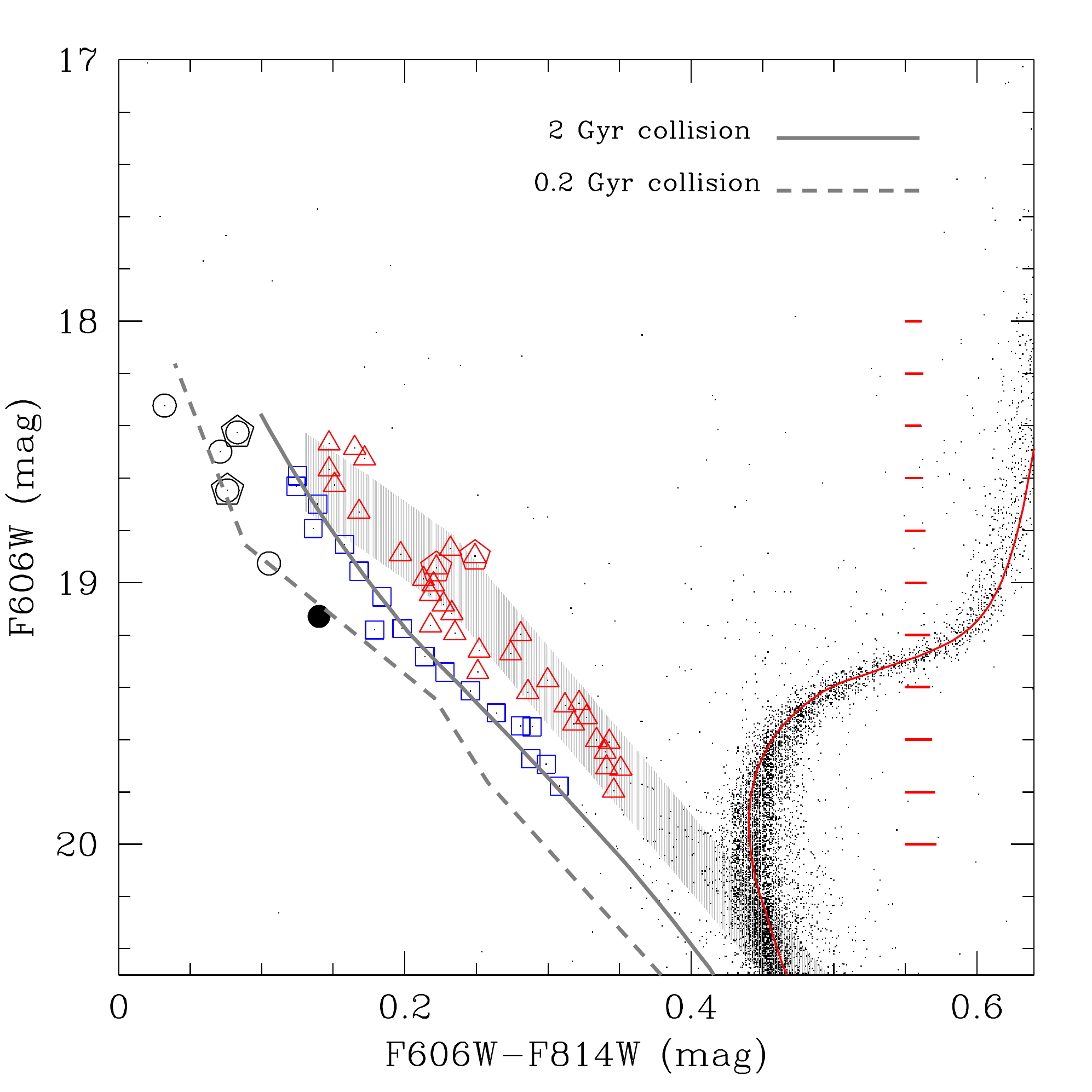}
\caption{\small{(F606W-F814W)\,vs.\,F606W CMD of the NGC\,1261 inner region with overplotted collisional isochrones of 2 Gyr and 200 Myr old, up to 1.3 $M_\odot$ and 1.6 $M_\odot$, respectively.~The grey band shows the zero-age main sequence isochrone, shifted by 0.45 and 0.75\,mag to brighter luminosities, to match the R-BSS sequence.~The red line is a Dartmouth isochrone with cluster parameters adopted from \cite{dot10}.~The representative photometric errors as obtained in the HST/ACS catalog are plotted in red bars.} }
\label{cmd_iso}
\end{figure}

We now introduce the F336W-band photometry for contaminant detection.~We show in Figure~\ref{Ucmd} the (F336W-F814W)\,vs.\,F336W CMD and use the same symbols for the BSSs as in Figure~\ref{cmd1}.~Three BSS candidates plus another star from the eB-BSS group cannot be distinguished from $normal$ MS/SGB stars based on their location in the (F336W-F814W)\,vs.\,F336W CMD (shown as diagonal crosses and a filled circle, respectively), and hence we exclude them from the full BSS sample and subsequent analysis.~It is worth noting that two R-BSSs and another two eB-BSSs in Figure~\ref{Ucmd} (additionally marked with a pentagon) also show significantly redder (F336W-F814W) colors than the rest of their groups.~This result may point into photometric variability (since the F336W and F814W-band images on HST were taken $\sim\!7$ years apart, while the F606W and F814W-band exposures are less than an hour apart).~Indeed W\,UMa eclipsing binary systems are frequent among BSSs and have been detected in the double BSS sequences of NGC 362 (D13) and M\,30 (F09).~In particular, the W\,UMa BSSs detected in NGC\,362 were found to show a typical variability of 0.3 mag in all F390W, F555W and F814W bands (see Figure~10 in D13) which could account for the deviations found in (F336W-F814W) colors for these BSSs in NGC\,1261.~Another explanation could be the blend of cooler stars, which would explain the apparent missing flux in the F336W band.~However, the fact that the source centroids were requested to match within 0.02\arcsec\ ($\sim$0.2 FWHM in the optical) in the optical and near-ultraviolet, along with the conservative 0.03 mag error limit in the optical bands, points rather towards a well-fitted single PSF detection.~We compare now the identified sub-samples with isochrones from stellar collisional models of \cite{sil09}.~The models have a metallicity of $Z\!=\!0.001$ ([Fe/H]~$\!=\!-1.27$) and solar-scaled chemical composition ([$\alpha$/Fe]~$\!=\!0$).~Pairs of stars with masses between 0.4 and 0.8\,$M_\odot$ are collided using the MMAS software package \citep{lom02}.~The parent stars are assumed to be non-rotating, and 10 Gyr old at the time of the collision.~The collision products are evolved using the Monash stellar evolution code, as described in \cite{sil09}.~The time of the collision was taken to be $t\!=\!0$, and isochrones of various ages were calculated by interpolating along the tracks to determine the stellar properties (effective temperature, luminosity, etc.) at ages between 0.2 and 5 Gyr.~For the isochrones, we used collision products of the following stellar mass combinations: $0.4\!+\!0.4, 0.4\!+\!0.5, 0.4\!+\!0.6, 0.5\!+\!0.6, 0.6\!+\!0.6$, and $0.8\!+\!0.8\,M_\odot$.~We have adopted the distance modulus and reddening values of NGC\,1261 from \cite{dot10} and augmented the distance modulus by 0.25 mag in order to match the low-mass end of the collisional isochrone to the location of the main-sequence fiducial line in the CMD\footnote{This is mostly caused by the different metallicity/alpha enhancement between our models and the cluster, which has a metallicity closer to [Fe/H]$\sim-$1.35...$-$1.38 according to the latest studies \citep{kra10,pau10,dot10}, and a large $\alpha$-enhancement as it is expected for GCs in the halo \citep{pritzl05,woodley10} .}.~We find that the location of the B-BSSs and eB-BSSs can be reproduced by 2 Gyr and 200 Myr old isochrones of the stellar collisional models, respectively, as seen in Figure~\ref{cmd_iso}.~We note that the agreement of the model with the B-BSSs is remarkably good.~The eB-BSS sub-sample, although much less populated, is also somewhat consistent with the collisional 0.2\,Gyr isochrone model.~This enlarges the likelihood of this BSS feature being a real distinct population.~The R-BSSs need further explanation.~F09 and D13 found for M\,30 and NGC\,362 that the lower bound of the R-BSSs could be fairly well bracketed by the zero-age main-sequence (ZAMS) shifted by 0.75 mag towards brighter luminosities in the $V$ (F555W in their case) band, which approximately indicates the region populated by mass-transfer binaries, as predicted by \cite{tia06}.~In our case we use the 0.25 Gyr old Dartmouth isochrones \citep{dot08}, and find that the R-BSS sequence can only be reproduced by a region bracketed by the 0.25 Gyr old isochrone shifted by 0.45 mag and 0.75 mag to brighter F606W-band luminosities (grey region in the Figure). This shift is less than that required for M\,30 and NGC\,362 , and we note that a small difference is expected due to slightly different HST filters (F606W vs. F555W) as a $V$ band proxy. However, the prediction by \citeauthor{tia06} is based only on Case A (main-sequence donor) mass-transfer models.~\cite{lu10} showed that case B (red-giant donor) mass-transfer products lie indeed in a bluer region than the ZAMS+0.75 mag boundary.~Our current understanding of binary mass transfer is limited and, hence, our observations could help putting constraints on future binary stellar evolution models. 

\section{Dynamical State of NGC\,1261}
\begin{figure}[t!]
\centering
\includegraphics[width=9cm]{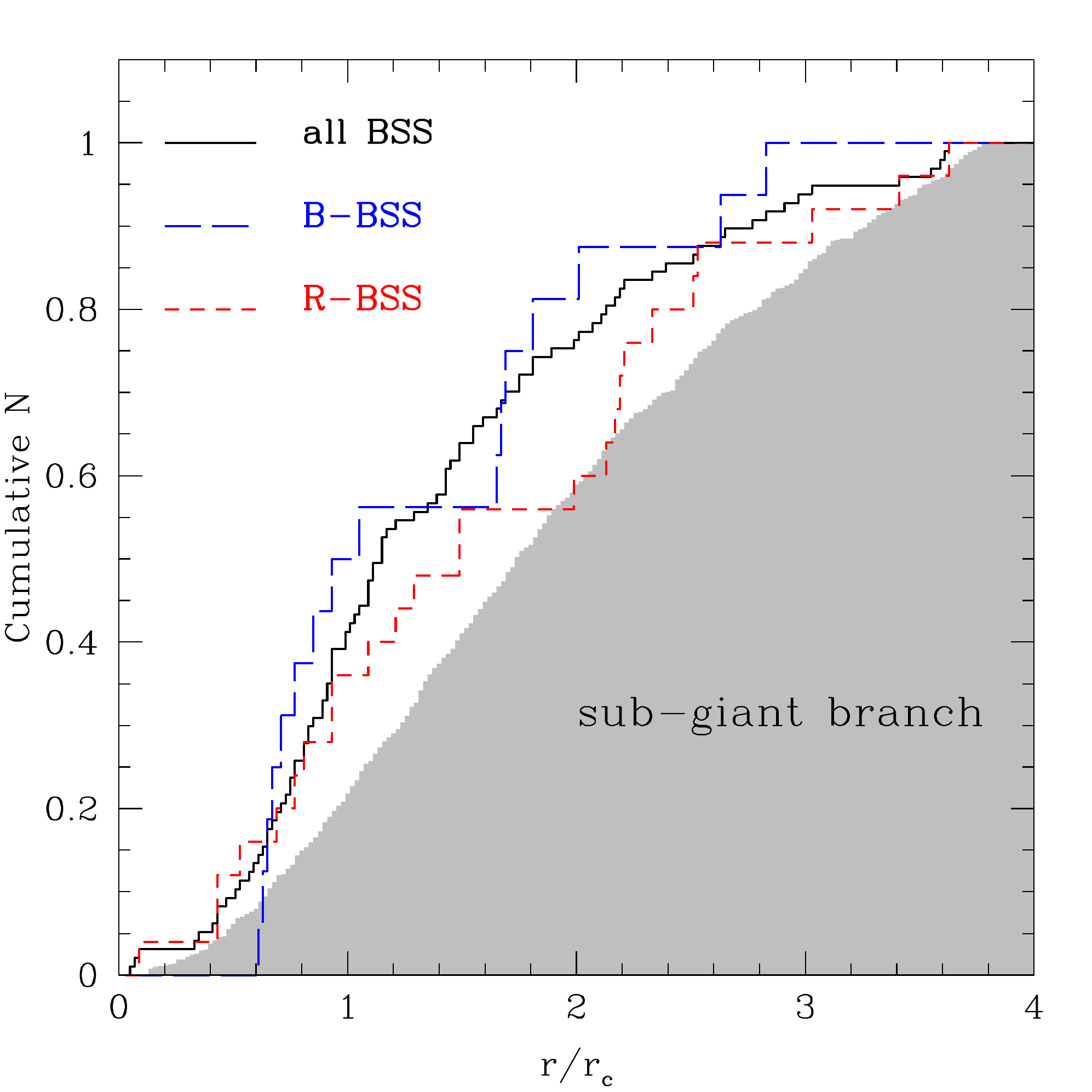}
\caption{\small{Normalized cumulative radial distribution of BSSs in the central 3.8$\,r_c$ of NGC\,1261.~The full BSS sample (solid black line), B-BSSs (long-dashed blue line) and R-BSSs (short-dashed red line) are plotted.~The corresponding cumulative distribution of the reference SGB population is shown as the shaded region.~We use the centre of gravity coordinates RA~$\!=\!$~03h\,12m\,16.21s, DEC~$\!=\!-55^{\rm o}\,12\arcmin\,58.4\arcsec$ given by \cite{gol10}, and $r_c=0.35'$ \citep{har96}.}}
\label{BSS_norm_prof}
\end{figure}

F09 and D13 have demonstrated that M\,30 and NGC\,362 show signs of being in an advanced state of dynamical evolution, as revealed by their centrally segregated BSS radial profiles \citep[which puts them in the {\sc Family~III} group in][]{fer12} and by their centrally peaked radial stellar density profiles.~Likewise, we plot in Figure~\ref{BSS_norm_prof} the normalized, cumulative BSS radial density profile for all BSS candidates and for the R-BSS and B-BSS sequences in NGC\,1261 out to 3.8 core radii, i.e.~the extent of the HST observations.~As expected, the whole BSS sample, as well as each BSS sequence, are more centrally concentrated than the reference SGB population\footnote{We choose SGB stars from the CMD inside the 19.2$<$F606W$<$19.5 mag range.}.~However, contrary to what was found in M\,30 and NGC\,362, we find the B-BSS component more centrally concentrated than the R-BSS component.~A K-S test shows that the null-hypothesis of these stars being drawn from the same radial distribution has a p-value of 0.33, which implies a non-negligible likelihood for common parent distributions.~Nevertheless, the difference between these profiles and the ones for M\,30 and NGC\,362 is still significant and motivates a discussion on possible qualitative differences of BSS formation history in NGC\,1261 versus the other GCs.~We do not find any B-BSSs inside $\sim\!0.5\,r_c$ (or about 10\arcsec), in agreement with F09 and D13, who as well report no B-BSSs within the inner 5-6\arcsec ($\sim\!1.5\,r_c$ and $\sim\!0.5\,r_c$ respectively).~These authors suggest that dynamical kicks are responsible for clearing the innermost region of any B-BSSs.~The detailed process is, however, unknown as accurate dynamical models are numerically expensive and, thus, still lacking.

The results from Figure~\ref{BSS_norm_prof} alone cannot be used to suggest an advanced dynamical state in NGC\,1261, as the central concentration of BSSs is now known to be ubiquitous among all studied GCs.~A more complete understanding can be obtained by looking at the BSS radial profile at larger cluster-centric distances, as the radial distance of the normalized BSS density minimum will depend on the cluster's dynamical age.~\cite{fer12} showed that as a GC evolves dynamically the BSS radial density profile takes the form of a central high maximum and a secondary peak at larger radii with a minimum in between. In order to follow this approach we include the wide-field photometry catalog in our analysis and carefully merge it \footnote{The resulting merged SGB radial profile distribution is checked to be smooth and continuous, therefore, ruling out severe completeness issues for the BSSs, which are in a similar magnitude range.~The merging point is chosen at 3.8 $r_c$.} with the inner-region ACS catalog, providing us with a sampling of NGC\,1261 out to $r\!>\!30\,r_c$. 

\begin{figure}[t!]
\centering
\includegraphics[width=9cm]{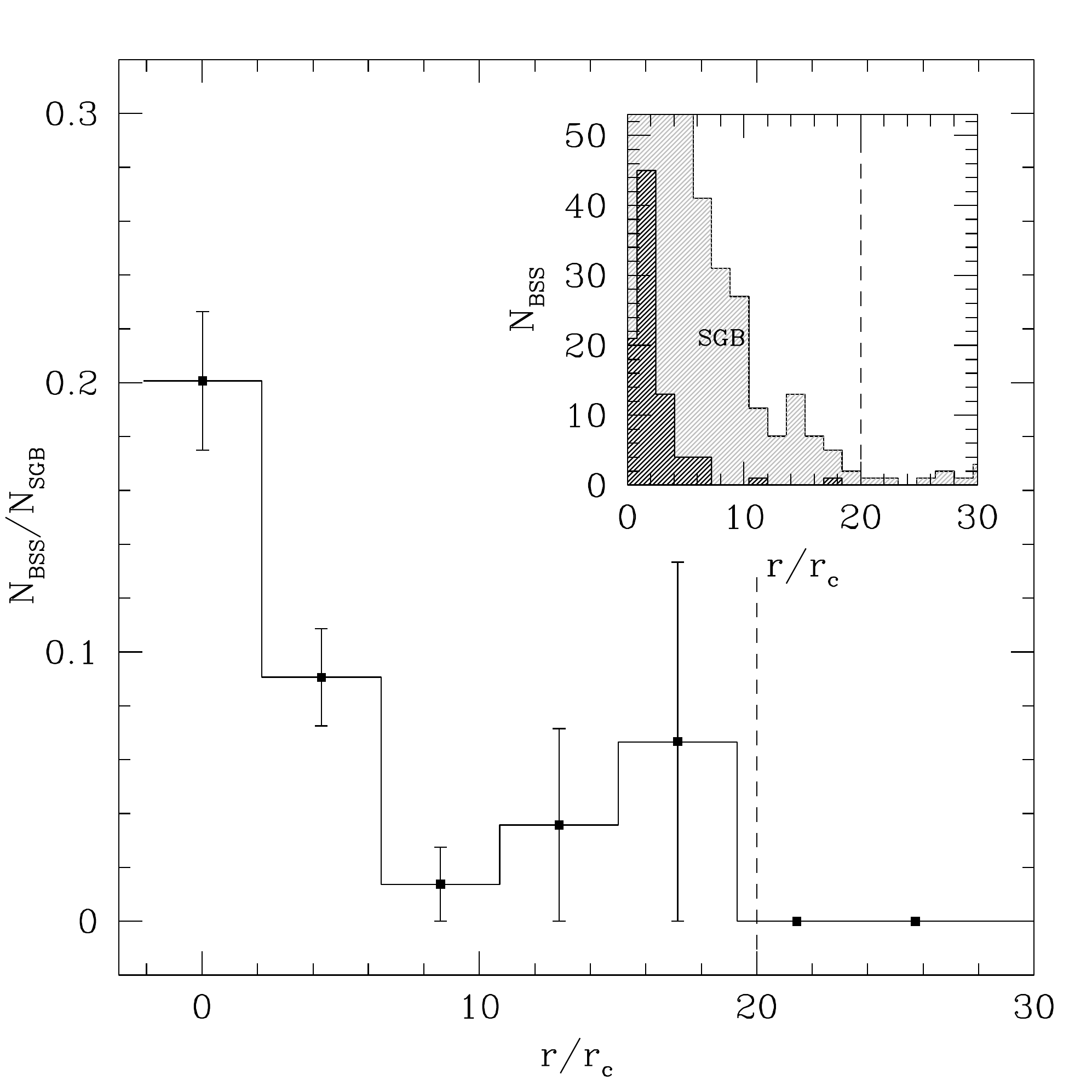}
\caption{\small{Ratio of BSS to SGB stars as a function of radial distance.~The inset panel shows the number of BSSs and SGB stars. Both panels have a dashed vertical line indicating the approximate location of the cluster's tidal radius, according to \cite{pau10}}.}
\label{BSS_rad_frac}
\end{figure}

We plot in Figure~\ref{BSS_rad_frac} the relative fraction of BSSs to SGBs\footnote{They were selected using a $19.2\!<\!V\!<\!19.5$ mag range cut. We use the ``$V$ $ground$" magnitude on the ACS catalog in order to make them compatible with the wide-field gound-based photometry.} as a function of cluster-centric radius, assuming Poissonian noise for the error bars.~The BSSs in the wide-field catalog are the ones selected by \cite{kra10}, i.e. stars on the BSS region of all ($B$-$V$), ($V$-$I$) and ($B$-$I$) CMDs.~We then apply an additional magnitude cut ($I<19.45$ mag) in order to hold the same faint magnitude limit used with the inner sample (see Fig~\ref{cmd1})\footnote{The F814W$<$19.5 mag limit from Fig.~1 is checked using the ``$I$ $ground$" magnitude on the ACS catalog and is found to translate to $I < 19.45$ mag.}.~Considering these selection steps, combining both samples is hence acceptable for our purposes.~We find that the BSS fraction is maximal in the center and drops rapidly with radius reaching near to zero at $r\!\sim\!10\,r_c$.~A subsequent rising in the fraction profile is discernible, although this is only caused by the detection of two BSSs alone (see the inset panel) and therefore not strongly supported statistically.~The absence of a clear outer layer suggests that the majority of the BSS population has already been affected by dynamical friction. Hence, in the framework constructed by \cite{fer12}, NGC\,1261 would classify as a dynamically old cluster and would be grouped in late-{\sc Family\,II}/{\sc Family\,III} , not surprisingly similar to M\,30 and NGC\,362.~Moreover, the half-mass relaxation time is about $10^8$ years, which is actually shorter than that of NGC\,362, $\log t=8.7$ and M\,30, $\log t=9.2$, as found in \cite{pau10}. 

However, the core structure of NGC 1261 does not show signatures of core collapse.~It is well approximated by a King model with a concentration parameter $c\!\approx\!1.2$ \citep{har96, pau10}.~It has a central luminosity density of 2.22 $L_{\odot}$pc$^{-3}$ \citep{pau10}, which is relatively low for GCs.~Moreover, the binary fraction radial profile found by \cite{mil12} shows a flat distribution, i.e.~without signs of mass segregation, contrary to the ones in M\,30 and NGC\,362, which are centrally peaked.~Therefore it is not surprising that the binary merger products, i.e.~the R-BSS population, in these GCs are also more centrally segregated than the R-BSS population in NGC\,1261.~This apparent evolutionary contradiction may not be as problematic as it first seems.~According to both Monte-Carlo dynamical models \citep[e.g.][]{heg08} and direct-integration N-body models \citep[e.g.][]{hur12}, clusters can go through core-collapse and then, if there is some energy source in the core, stay or pass through a long-lived post-core-collapse bounce state in which they do not show classic post core-collapse signatures.~The \citeauthor{heg08} models of M\,4 all went through core collapse at $t\!=\!8$\,Gyr, and then remained in a non-collapsed state for another $\sim\!2\!-\!3$ Gyr due to ``binary burning" in the core. \citeauthor{hur12} propose a binary black hole as an alternative central potential energy source.~We know today that black holes may be common in GCs \citep[e.g.][]{str12, cho13}, so that a binary black hole in the core of NGC\,1261 may be considered a valid possibility.\\
 
\section{Summary and Conclusions}
We find that the inner BSS population in NGC\,1261 includes at least two distinct well defined sequences similar to what was found in M\,30 and NGC\,362, and as well includes a smaller group of BSSs that have unusually blue colours in the CMD, and which could be associated with a distinct coeval BSS population, if confirmed real.~The comparison with collisional stellar evolution models reveals that the B-BSS and eB-BSS sub-samples are consistent with a 2 Gyr and 0.2 Gyr old stellar collision-product population, respectively. This provides the grounds for considering NGC\,1261 an extremely valuable test laboratory for stellar collision and BSS formation models.~This observation along with evidence collected from the literature suggest as a preliminary interpretation for the dynamical history of NGC\,1261 that the cluster experienced a core-collapse phase about 2 Gyr ago, and that since then it has bounced through core oscillations.~The subsequent core oscillations occasionally created more BSSs during these short timescale processes when the central stellar density was particularly enhanced -- one such likely 0.2 Gyr ago. During these periods of core density enhancements the cluster likely burned some of its core binaries, thereby flattening their radial density distribution profile, and currently the cluster is likely in a post core-collapsed state which, according to different simulations, may appear as an unevolved GC. Follow-up spectroscopic characterization of the BSS sequences in NGC\,1261 is of utmost importance in order to confirm and better understand their origins and formation mechanisms, and in particular test the chemical abundance predictions related to different BSS formation models.

\acknowledgments
We thank the anonymous referee for comments that improved the presentation and quality of our results.~MS and THP gratefully acknowledge support from CONICYT through the ALMA-CONICYT Project No.~37070887, FONDECYT Regular Project No.~1121005, FONDAP Center for Astrophysics (15010003), and BASAL Center for Astrophysics and Associated Technologies (PFB-06), as well as support from {\it Deutscher Akademischer Austauschdienst} (DAAD). AS is supported by NSERC.~This research has made use of the Aladin plot tool and the TOPCAT table manipulation software, found at http://www.starlink.ac.uk/topcat/.
{\it Facilities:} \facility{HST (ACS)}.

\clearpage

\end{document}